\documentclass[]{spie}

\usepackage[]{graphicx}
\usepackage{amsmath, amssymb, amsfonts}
\usepackage[english]{babel}

\usepackage{booktabs,caption,fixltx2e}
\usepackage[flushleft]{threeparttable}
\usepackage[font={small,it}, margin = 0.5cm]{caption}

\title{Correcting for the effects of pupil discontinuities with the ACAD method} 

\author{Johan Mazoyer\supit{a}, Laurent Pueyo\supit{a}, Mamadou N'Diaye\supit{a}, Dimitri Mawet\supit{b}, R\'e{}mi Soummer\supit{a}, Colin Norman\supit{c}
\skiplinehalf
\supit{a} Space telescope Science Institute, 3700 San Martin Drive, Baltimore, MD 21218, USA\\
\supit{b} Astronomy Department, Caltech University, Pasadena, CA, USA\\
\supit{c} Department of Physics and Astronomy, Johns Hopkins University, Baltimore, MD, USA\\
}

\authorinfo{Further author information: contact Johan Mazoyer at jmazoyer@stsci.edu}
 
\begin{document} 
  
\maketitle 

\begin{abstract}
The current generation of ground-based coronagraphic instruments uses deformable mirrors to correct for phase errors and to improve contrast levels at small angular separations. Improving these techniques, several space and ground based instruments are currently developed using two deformable mirrors to correct for both phase and amplitude errors. However, as wavefront control techniques improve, more complex telescope pupil geometries (support structures, segmentation) will soon be a limiting factor for these next generation coronagraphic instruments. The technique presented in this proceeding, the Active Correction of Aperture Discontinuities method, is taking advantage of the fact that most future coronagraphic instruments will include two deformable mirrors, and is proposing to find the shapes and actuator movements to correct for the effect introduced by these complex pupil geometries. For any coronagraph previously designed for continuous apertures, this technique allow to obtain similar performance in contrast with a complex aperture (with segmented and secondary mirror support structures), with high throughput and flexibility to adapt to changing pupil geometry (e.g. in case of segment failure or maintenance of the segments).

We here present the results of the parametric analysis realized on the WFIRST pupil for which we obtained high contrast levels with several deformable mirror setups (size, separation between them), coronagraphs (Vortex charge 2, vortex charge 4, APLC) and spectral bandwidths. However, because contrast levels and separation are not the only metrics to maximize the scientific return of an instrument, we also included in this study the influence of these deformable mirror shapes on the throughput of the instrument and sensitivity to pointing jitters. Finally, we present results obtained on another potential space based telescope segmented aperture.

The main result of this proceeding is that we now obtain comparable performance than the coronagraphs previously designed for WFIRST. First result from the parametric analysis strongly suggest that the 2 deformable mirror set up (size and distance between them) have a important impact on the performance in contrast and throughput of the final instrument. 

\end{abstract}



\section{Introduction}
\label{sec:Intro}

Several coronagraph designs were developed over the last decade for the current generation of high contrast ground based instruments \cite{Beuzit08,Macintosh08,Hinkley11} mostly designed for circular, unobstructed pupils. In some cases (Apodized Pupil Lyot Coronagraph, APLC \cite{Soummer11}, or ring apodized vortex \cite{Mawet13}{}), they may have taken into account the central obscuration. This generation of coronagraph designs was able to reach the target level of contrast ($\sim 10^{-5}$). However, the new goal in the quest for the highest contrast levels is now the correction of the diffractive effects introduced in the focal plane of coronagraph by discontinuities in the apertures. 

The next generation of space based telescopes (WFIRST \cite{Spergel2015}{}, ATLAST \cite{Feinberg2014} or HDST\cite{Dalcanton2015}) and of ground based coronagraphic instruments (PCS for E-ELT\cite{Kasper10}{}, or the TMT\cite{Macintosh06}) will be mounted on on-axis and/or segmented telescopes. To reach the desired levels of contrast of ($\sim 10^{-10}$ for space based instruments, $\sim 10^{-8}$ for ground based instruments) the design of coronagraphic instruments for such apertures is currently a domain undergoing rapid progress. Specific coronagraphic designs (phase- induced amplitude apodization with complex mask coronagraph\cite{Guyon14}{}, shaped pupil coronagraph\cite{Carlotti14} or APLC \cite{Soummer05, Ndiaye16}) are currently developed to reach high contrast levels after such pupils. These techniques will be used in addition with deformable mirrors (DMs) to correct for wavefront errors, either residuals of the adaptive optic system on ground based telescopes, or introduced by the optics themselves for space and ground based telescopes. The current designs for future high-contrast instruments now systematically include two sequential DMs for the simultaneous correction of phase and amplitude wavefront errors on a symmetrical dark hole.

However, several coronagraphs \cite{Kuchner02,Guyon05,Mawet13, Soummer11, Ndiaye15} have already been developed for circular axisymmetric apertures, with various inner working angle and contrast performance. Another approach therefore consists of using these coronagraphs and the possibilities of the already provided sequential DMs to correct specifically for aberrations introduced by secondary mirror structures and segmentation of the primary mirror. The method used in this paper, Active Correction of Aperture Discontinuities (ACAD), originally introduced by Pueyo \& Norman (2013)\cite{Pueyo_Norman13}{}, was developed specifically for that goal. It aims at returning to the performance in contrast level and inner working angle (IWA) obtained on axisymmetric apertures with various coronagraphs. However, recent studies \cite{Stark15} emphasis the importance of other metrics (throughput, robustness to jitter, spectral bandwidth) in the yield of exoplanet detected by future missions, which need to be also taken into account in these studies. The analysis presented here studies the influence of several parameters (form of the pupil, size of the DMs and distance between them, type of coronagraph, spectral bandwidth) on these metrics (contrast level, Throughput, robustness to jitter). We hope that a better understanding of the effects of these parameters on the performance of the system will help constrain the design of future high contrast instruments.

In Section \ref{sec:acad}, we recall the method and the recent optimisations we developed. We then describe the parameter space in Section~\ref{sec:parameter_space}, and present the results of the parametric analysis in Section~\ref{sec:wfirst}. Finally, we will use the ACAD technique on a realistic apertures for future space based missions in Section~\ref{sec:SCDA}

\section{Description of the ACAD method}
\label{sec:acad}

\begin{figure}[ht]
 \begin{center}
    \includegraphics[width = 0.7\textwidth, clip = true]{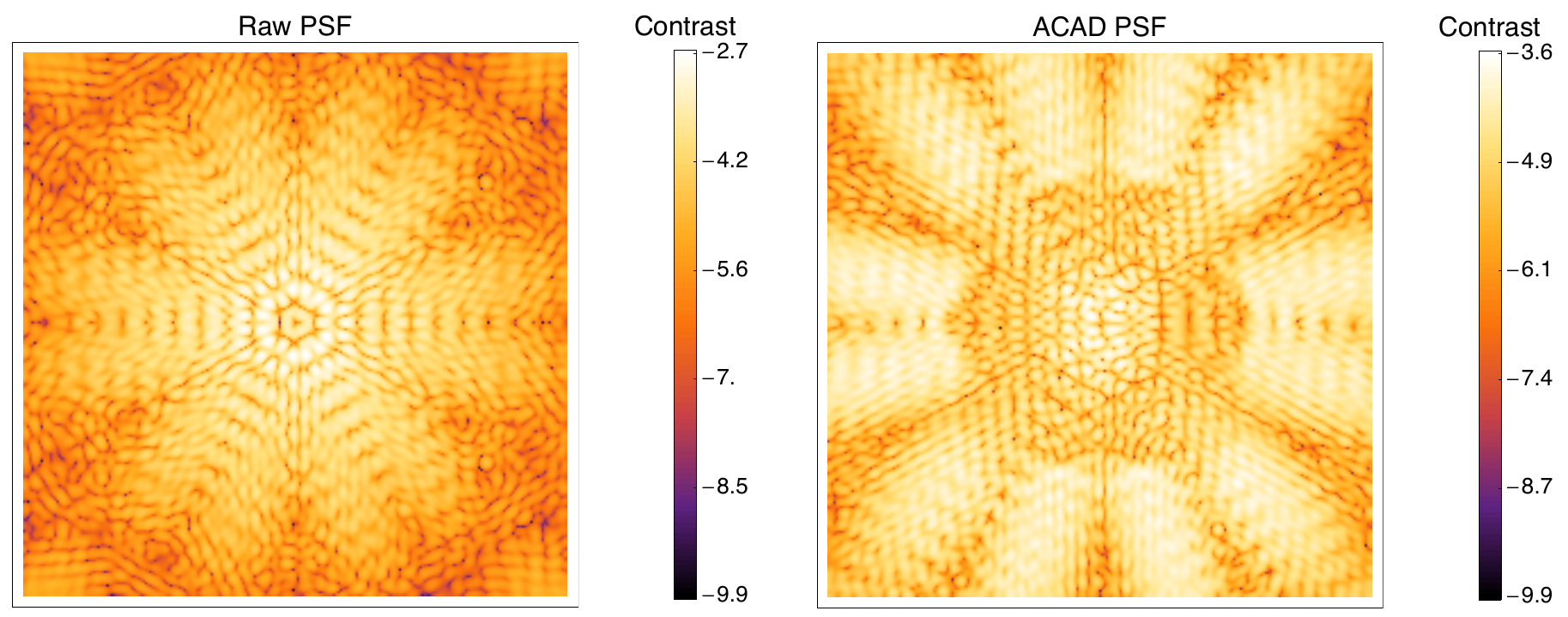}
\end{center}
 \caption[] 
{ \label{fig:psf_acad} 
Two PSFs, after the WFIRST pupil and a vortex coronagraph. Left: PSF with flat mirrors, showing the diffraction effects of the pupil discontinuities. Right: same PSF after applying geometrical Monge-Amp\`ere solution}
\end{figure}

In previous papers, (Pueyo \& Norman 2013\cite{Pueyo_Norman13}{}, Pueyo et al. 2014\cite{Pueyo14}{}, Mazoyer et al. 2015\cite{Mazoyer_SPIE15}), the ACAD method was presented in two steps. The first step is an analytical ray optic solution for the DM shapes obtained by the resolution of the Monge-Amp\`ere equation. Figure~\ref{fig:psf_acad} shows the PSF obtained in the focal plane of a vortex coronagraph, after the apertures of the WFIRST telescope. On the left, before the correction, the DM are both flat, and the PSF clearly shows the diffraction effects of the pupil discontinuities. On the right, we applied the analytical shapes solution obtained by the resolution of the Monge-Amp\`ere equation on the 2 DMs. The strokes introduced by this first step of the correction is strongly dependent on the DM size and inter-DM distance but it easily varies from a few hundreds of nanometer to a few micrometers.
In the correction zone of the mirror, the diffraction effects of the pupil discontinuities are mitigated: the DM shapes obtained after this ray optic solution are clearly remapping the electrical field to obtain a flat wavefront in the entrance of the coronagraph. However, the contrast level in this zone have barely improve. This is not surprising and had been observed multiple time \cite{Borde06, Mazoyer13}{}: \textit{a flat wavefront produces an excellent contrast at all separation, but in the context of extreme adaptive optics and DM with limited degrees of freedom, trying to flatten the wavefront is not best option to obtain the best contrast}. This is why, on top of the ray optic solution, we ran a stroke minimization algorithm (SM, Pueyo et al. 2009 \cite{Pueyo09}{}). This second step was producing the zone of high contrast in the focal plane of the coronagraph, called dark hole (DH). 
\begin{figure}[ht]
 \begin{center}
    \includegraphics[width = 0.5\textwidth,trim= 0cm 0.5cm 0.5cm 0.3cm, clip = true]{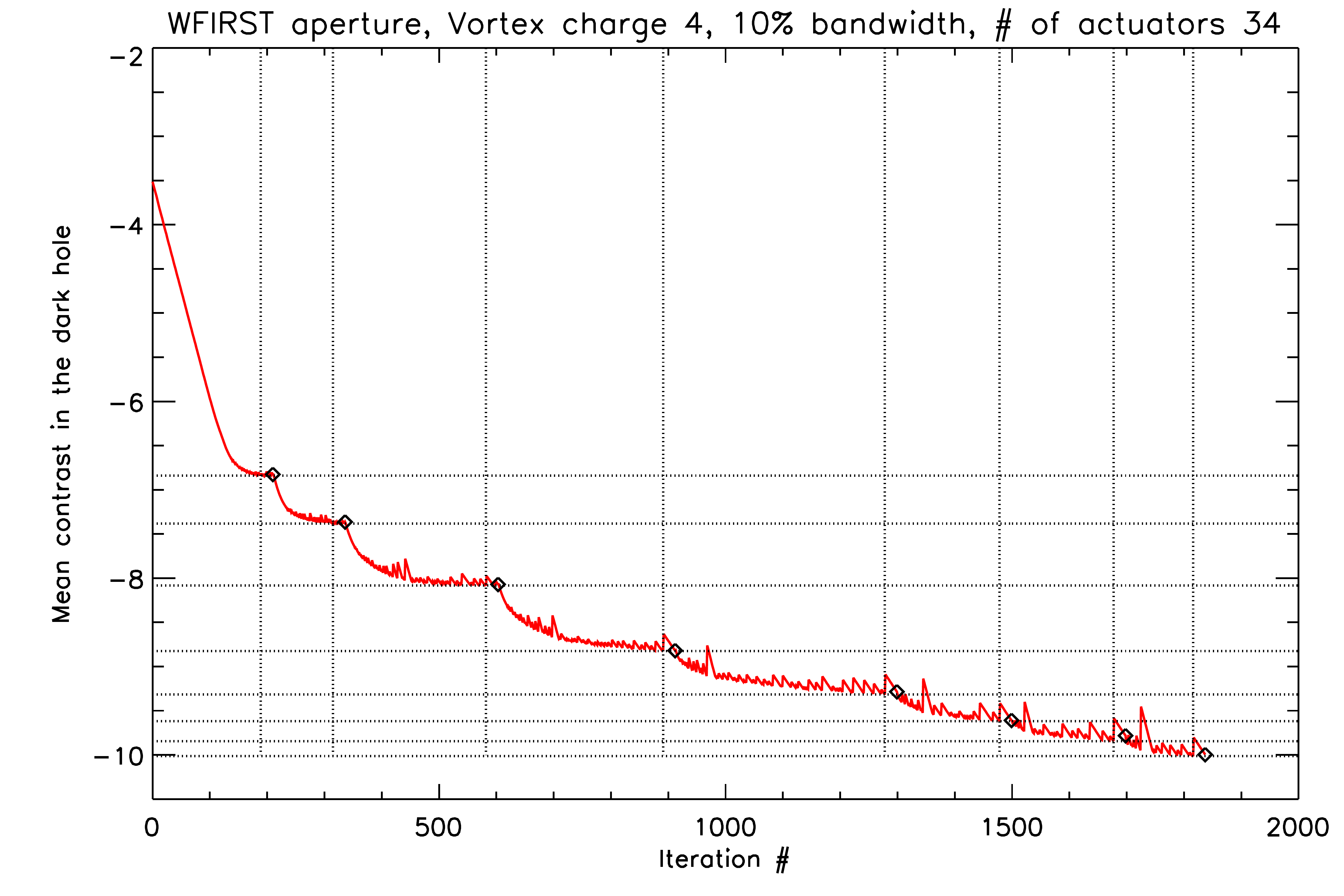}
\end{center}
 \caption[] 
{ \label{fig:iter_strokemin} 
Mean ontrast level in the DH as a function of the number of iterations, for 8 matrices of the SM algorithm. The dotted vertical and horizontal lines correspond to the best contrast reached for each matrix and the iteration for which it was reached.}
\end{figure}

However, the performance of these combined two steps were disappointing: the final contrast levels in the DH was never better than $10^{-8}$ for a 10\% bandwidth. This was due to the fact that, contrarily to the ray optic solution, the SM is a linear correction algorithm. The linear approximation is only valid if the phase introduced by the DM is small enough. We realized that the SM algorithm was always diverging at some point, usually when the introduce strokes were higher than $\sim \lambda/5$. This problem of the SM algorithm had not arisen in previous uses\cite{Pueyo09} because it is usually used to correct relatively small aberrations in continuous apertures. To solve this problem, we first optimized the SM algorithm with a gain changing when the algorithm starts to diverge. We then identity the moment the algorithm is definitely diverging and recalculated the interaction matrix around this point. This ensures that the strokes stay limited under $\lambda/5$ around each new initial DM positions.

Fig~\ref{fig:iter_strokemin} shows the contrast obtained as a function of the number of iterations for a SM algorithm with 10\% bandwidth, WFIRST pupil (charge 4 vortex, D = 34x0.3 mm, Z = 0.3m). Horizontal and vertical doted lines indicate, for every matrices, the best contrast level reached and the iteration number for which it was obtained. The DMs shapes to obtain this contrast level are saved. When SM algorithm is diverging for more than 20 iterations (indicated by black diamonds), we interrupt the loop and re-calculate the interaction matrix. On this figure, 8 matrices were necessary. We see that the improvement in contrast level for the last matrix is of a factor $10^{-9.84}/10^{-10.01}$ which is only a factor 1.5 (which meant that a ninth matrix would probably barely improve the contrast). For all of our corrections in broadband, we use 8 matrices. For monochromatic corrections, which are usually a lot faster to converge, we use 5 only. The final stroke introduced on the DMs by the SM algorithm are aperture dependant but are usually under 150nm, even for the WFIRST aperture. 

At that point we realized that this multi-matrix SM was so effective to obtain high contrast level in DH that starting from the ray optic solution or from a flat surface DM was giving comparable results in contrast level. However, the final DM shapes are very different: ray optic solution add important strokes, from hundreds of nanometers to a few micrometers to the final DM surfaces. High strokes pose several problems in the ACAD method. First, the DMs usually used for high contrast imaging (relatively small diameter, thousands of actuators) cannot currently reach stroke larger than a few hundreds nanometers (a few microns at most). Secondly and more importantly, high strokes have a strong impact on the out-of-axis PSF (this is the reason PIAA coronagraphs often need anti-PIAA to improve throughput), which strongly degrades the throughput.  

For these reasons, we chose in this proceeding to focus on the solution without the ray optic solution. However, it is possible that this latest solution has advantages, in particular for large bandwidths. This will be studied in an upcoming paper. 
However, even if it finally appear that the ray optic solution is not useful here, this method, which return the mirror shape necessary to obtain any wanted apodization, is still extremely powerful and useful for other applications.

\section{Explored parameter space}
\label{sec:parameter_space}

In this section, we explore the parameter space that have been explored in this study. We used 1 aperture (Section~\ref{sec:Apertures}), 3 DM set-ups (Section~\ref{sec:DMsetup}), 3 coronagraphs (Section~\ref{sec:corono_param}) and 3 bandwidths (Section~\ref{sec:BW}), which total to 27 different SM corrections. 

\subsection{Apertures}
\label{sec:Apertures}

The parametric analysis presented in Section~\ref{sec:wfirst} only uses the WFIRST pupil (Figure~\ref{fig:WFIRST_pup_foc}, top left). We are aware that the coronagraphic instrument designs have already been selected for this telescope \cite{Krist16}{}. However, the challenges presented by this pupil (large central obscuration and struts) are a good opportunity to put our method to the test. The goal of this parametric analysis is not to obtain performance comparable to the ones presented in Krist et al. 2015\cite{Krist16}{}, but to understand the relative influence of the studied parameters. For this reason, the WFIRST aperture is particularly adapted, allowing us to obtain a good range of contrast performance, from $10^{-7}$ to $10^{-12}$ depending on the coronagraph, bandwidth and DM setup selected. The radius of the central obstruction of the WFIRST pupil is 36\% of the radius. 

\subsection{Coronagraphy}
\label{sec:corono_param}

We use 3 different coronagraph designs in this study. The first one in an APLC, designed using the method described in N'Diaye et al. 2015\cite{Ndiaye15}{}. This coronagraph uses an optimized apodization associated to a classical Lyot mask is design to obtain a $10^{-9}$ contrast level, on a DH of $5-40 \lambda/D$ over a 10\% bandwidth for an axisymmetric pupil with a central obstruction of 36\%. New techniques of apodization optimization \cite{Ndiaye16} have since been developed and this coronagraph probably does not represent the best performance in contrast achievable with an APLC and the ACAD technique. For a large central obstruction (36 \%) the  APLC focal mask radius is important (5 $\lambda/D$). The Lyot stop include a central obstruction which radius represents 50\% of the entrance pupil radius. 

The last 2 coronagraphs are two ring apodized vortex coronagraphs (Mawet et al. 2005\cite{Mawet05}, Mawet et al, 2013\cite{Mawet13}), of charge 2 and 4. These coronagraphs are achromatic, and obtain analytically a perfect contrast (in absence of iterations) for an unaberrated apertures with a central obstruction of 36\%. We simulate the vortex coronagraph using the method described in Mazoyer et al. 2015\cite{Mazoyer_SPIE15}{}. The design of their axisymmetric apodization and of the radius of the Lyot stop central obstruction are analytically derived using equations in Mawet et al, 2013\cite{Mawet13}, and given in Table~\ref{tab:corono}.

To compare the performance of the coronagraphs, we decided to use a fixed inner working angle (IWA) and outer working angle (OWA) to create the DH. We choose to deepen $3-10 \lambda/D$ DH for the vortex coronagraph and $5-12 \lambda/D$ for the APLC.

\begin{table}[ht]
\centering
\begin{threeparttable}
\caption{Vortex coronagraphs parameters}
\label{tab:corono}
\begin{tabular}{lccc}
                                 & Charge 2                     & \multicolumn{2}{c}{Charge 4}              \\
                           \hline      \hline     
Pupil COR$^a$        & 0.36 R                                & 0.36 R          & 0.17 R                                             \\ \hline 
Ring apodization          & \begin{tabular}[c]{@{}c@{}}t=1 in 0.36 R\textless r \textless 0.71 R\\ t=0.67 in 0.67 R\textless r \textless R\end{tabular} & \begin{tabular}[c]{@{}c@{}}t=1 in 0.36 R\textless r \textless 0.69 R\\ t= 0 in 0.69 R\textless r \textless 0.78 R\\ t= 0.57 in 0.78 R\textless r \textless R\end{tabular} & \begin{tabular}[c]{@{}c@{}}t=1 in 0.36 R\textless r \textless 0.53 R\\ t= 0 in 0.53 R\textless r \textless 0.55 R\\ t= 0.81 in 0.55 R\textless r \textless R\end{tabular} \\
Lyot COR$^a$ & 0.67 R            & 0.78 R      & 0.55R           \\
\hline
\end{tabular}
	\begin{tablenotes}
      		\small
     		 \item[$^a$] Central obscuration radius (compared to the outer radius R of the pupil)
      	\end{tablenotes}
\end{threeparttable}
\end{table}

In the last column of Table~\ref{tab:corono}, we indicate the parameters for the charge 4 coronagraphs for 0.17\% central obscuration, that we are using in the simulation for the SCDA pupil that we present in Section~\ref{sec:SCDA}.

\subsection{Spectral bandwidth}
\label{sec:BW}

All the spectral bandwidths are given around a central wavelength of 550 nm. This central wavelength $\lambda_0$ is also used to defined the focal plane distance unit $\lambda_0/D$. We analyze three bandwidths in this study: 0\% (monochromatic case), 10\% and 20\%. The influence of spectral bandwidth on contrast level have been well studied before \cite{Shaklan06, Pueyo_Kasdin07}. Therefore, we are more particularly interested in this study to analyze its influence on the throughput and jitter in the context of the ACAD technique. For broadband SM algorithm, we use a multi-wavelentgh matrix: 3 wavelengths for 10\%, 5 wavelengths for 20\% and 7 for 30\%. However, once we obtain final DM shapes with SM algorithm, we propagate 20 wavelengths to produce focal planes and contrast curves. This allows us to check that our solutions are still correct with a more realistic spectral resolution. 

\begin{figure}[ht]
 \begin{center}
    \includegraphics[width = 0.49\textwidth, trim= 0cm 4.5cm 5cm 4cm, clip = true]{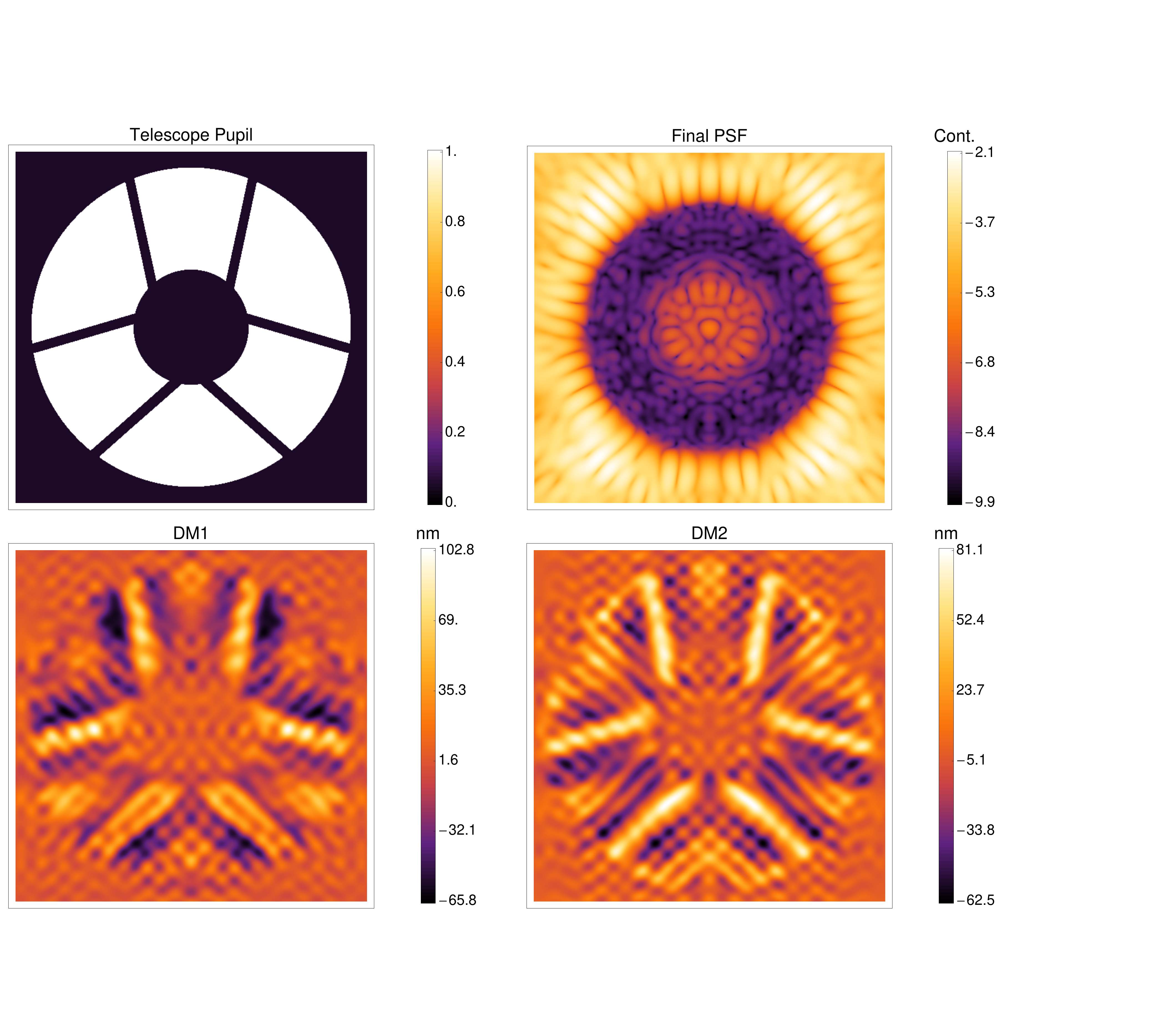}
    \includegraphics[width = 0.49\textwidth, trim= 0cm 4.5cm 5cm 4cm, clip = true]{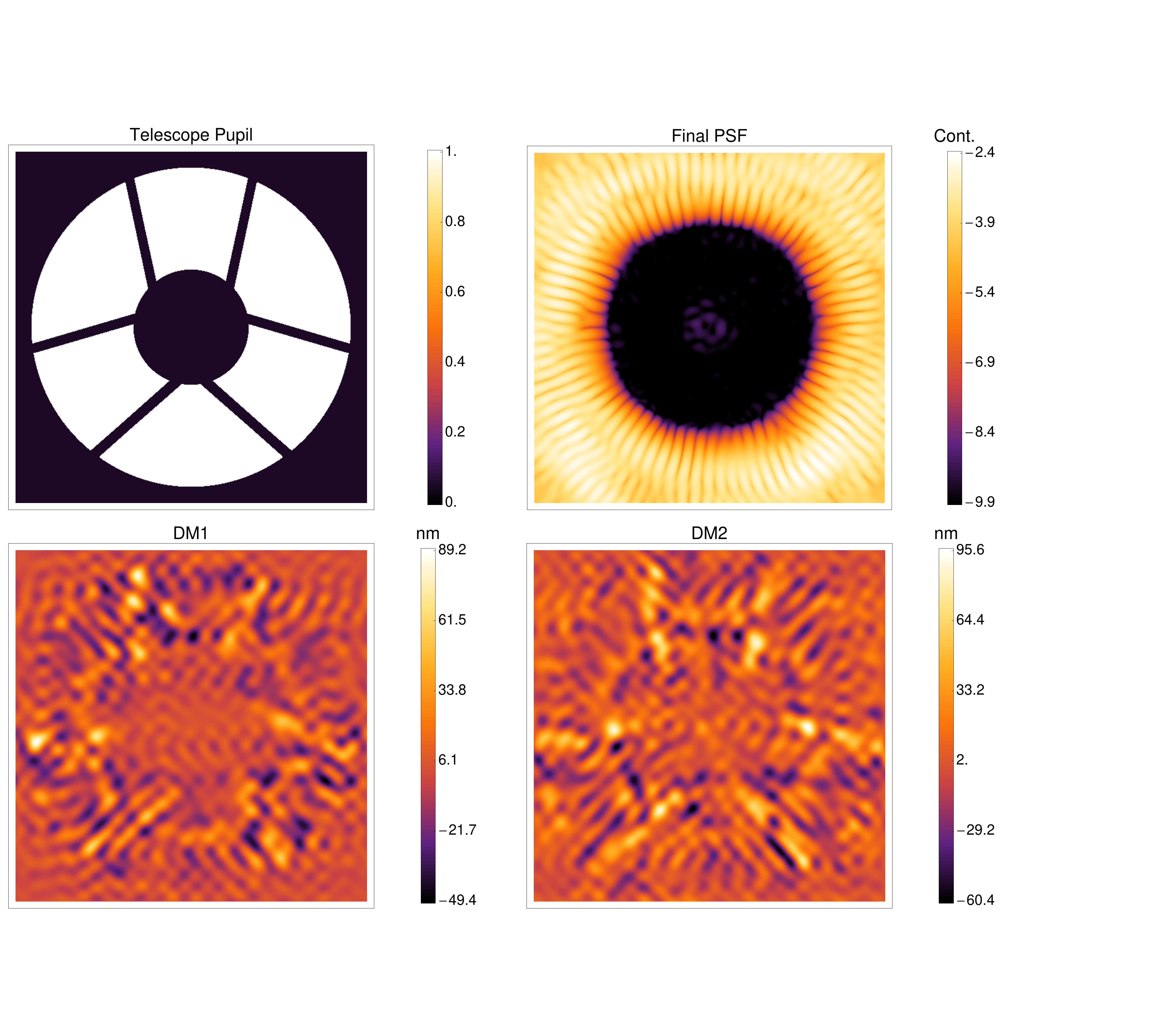}
\end{center}
 \caption[] 
{ \label{fig:WFIRST_pup_foc} 
Results obtain with ACAD on the WFIRST pupil, for the APLC (4 images on the left) and charge 4 vortex coronagraph (4 images on the right), with a 10\% bandwidth. The DMs have 34 actuators, a diameter of 34x0.3mm, and are separated by 30cm. For each coronagraph, we represent: the initial pupil (top left), the final DH obtained at the end of the correction (top right), 5-12 $\lambda$/D for APLC and 3 -10 $\lambda$/D for the Vortex, and the two shapes applied on the DMs to produce this DH.}
\end{figure}

\subsection{Deformable mirrors}
\label{sec:DMsetup}

We did not specifically study the influence of the number of actuators on the performance in this study. The influence of this parameter on the contrast level and outer working angle have already been studied in previous paper\cite{Borde06, Mazoyer13}. ACAD correction with a higher number of actuators can introduce higher spatial frequencies on the DMs, which will have then an impact on the throughput and robustness to jitter. That impact, that we think is going to be limited, will be studied later on this study. 

Since our goal in this section is to compare different configurations and not to obtain performance comparable to the ones presented in Krist et al. 2015\cite{Krist16}{}, we only used 34 actuators, inferior to the number of actuators planned for the WFIRST mission (48 actuator DMs). This number corresponds to  the number of actuators of the Boston Micromachines (BMC) DMs used on the HiCAT bench \cite{NDiayeSPIE15} and is small enough to be simulated quickly on a desktop computer. Because the number of actuators is fixed, we did not analyzed the influence of the OWA on the contrast. IWA and OWA have been selected to be comparable to the one used in Krist et al. 2015\cite{Krist16}{} (3 - 10 $\lambda/D$ for the vortex, 5 - 12 $\lambda/D$ for the APLC). The use of larger DMs would increase the performance in contrast on these DHs, or obtain comparable performance on larger DHs.

Finally, we analyzed 3 different DM setups, that we assume are representative of several experiments currently developed to analyze the segmentation problem. We study the two cases of "BMC" like DMs (inter-actuator distance of ~0.3 mm) and of "Xinetics" like DMs (inter-actuator distance of ~1 mm). This inter-actuator distance and the given number of actuators (34) constrain the size of the DMs to D = 1 cm for the BMC like DMs and D = 3.4 cm for the Xinetics like DMs. We then studied 2 inter-DM distances for the BMC like DMs (Z = 1 m and Z = 0.3 m) and one inter DM distance for the Xinetics like DM (Z = 1 m). The first setup (BMC DMs, D = 1 cm, and Z = 0.3 m) is the setup of the High-contrast imager for Complex Aperture Telescopes (HiCAT) bench\cite{NDiayeSPIE15}, on which the ACAD technique will be experimentally tested, and we will refer to it as HiCAT case. The second setup (Xinetics like DMs, D = 3.5 cm, and Z = 1 m) is the chosen setup for the WFIRST mission and we will refer to it as WFIRST case. We also test an "intermediate case" (BMC like DMs, D = 1 cm, and Z = 1 m). The parameters of these 3 different setups are reported in Table~\ref{tab:DMsetup}.

\begin{table}[ht]
\centering
\caption{Deformable Mirror setups in the parametric analysis}
\label{tab:DMsetup}
\begin{tabular}{l c c c}
                            & HiCAT case & WFIRST case & Intermediate case  \\   \hline  \hline    
\# of actuators              & 34         & 34             & 34               \\ 
DM Inter-act. distance & 0.3 mm  (BMC like)      & 1 mm (Xinetics like)         & 0.3 mm  (BMC like)            \\ 
Size of the DM (D)            & 1 cm         & 3.5 cm          & 1 cm                \\ 
Inter-DM distance (Z)      & 0.3 m        & 1 m              & 1 m                 \\ 
\end{tabular}
\end{table}

The pupil is slightly undersized compare to the DM. This is due to the fact that during the propagation between the two DMs, the beam slightly expands. This reduces the actual number of actuators in the pupil, leaving a part of the first DM outside of the pupil. However, this is often necessary as the imaged pupil on the second DM plane is larger than the actual pupil. If we take a pupil of the same size than the DMs, the propagated electrical field in the second DM plane might end up larger than the DM itself which would degrade the performance in contrast. This is particularly important if we use the geometrical approach of the ACAD method, where the actuator strokes are important, which can create important discontinuities at the edge of the second DM. In this study, we used a 10\% undersizing, which reduce the number of actuators in the diameter of the pupil to 31 (out of the 34 in the DM), which put a maximum limit to OWA to 15.5 $\lambda/D$. For the moment, we assumed that the surface around the second DM is perfectly reflective, but not deformable. We will also study the possibility of a non reflective material outside second DM, which will have a strong impact on throughput. In that case, we might have to increase the undersizing.

\section{Parametric analysis on the WFIRST pupil}
\label{sec:wfirst}
In this section, we analyse the performance of the 27 cases and present the results of the parametric analysis on the WFIRST aperture, in term of contrast level (Section~\ref{sec:contrast}), and throughput and robustness to jitter (Section~\ref{sec:Throughput}).

\subsection{Contrast levels}
\label{sec:contrast}
\begin{figure}[ht]
 \begin{center}
    \includegraphics[width = 0.495\textwidth, trim= 0.5cm 0cm 0.3cm 0cm, clip = true]{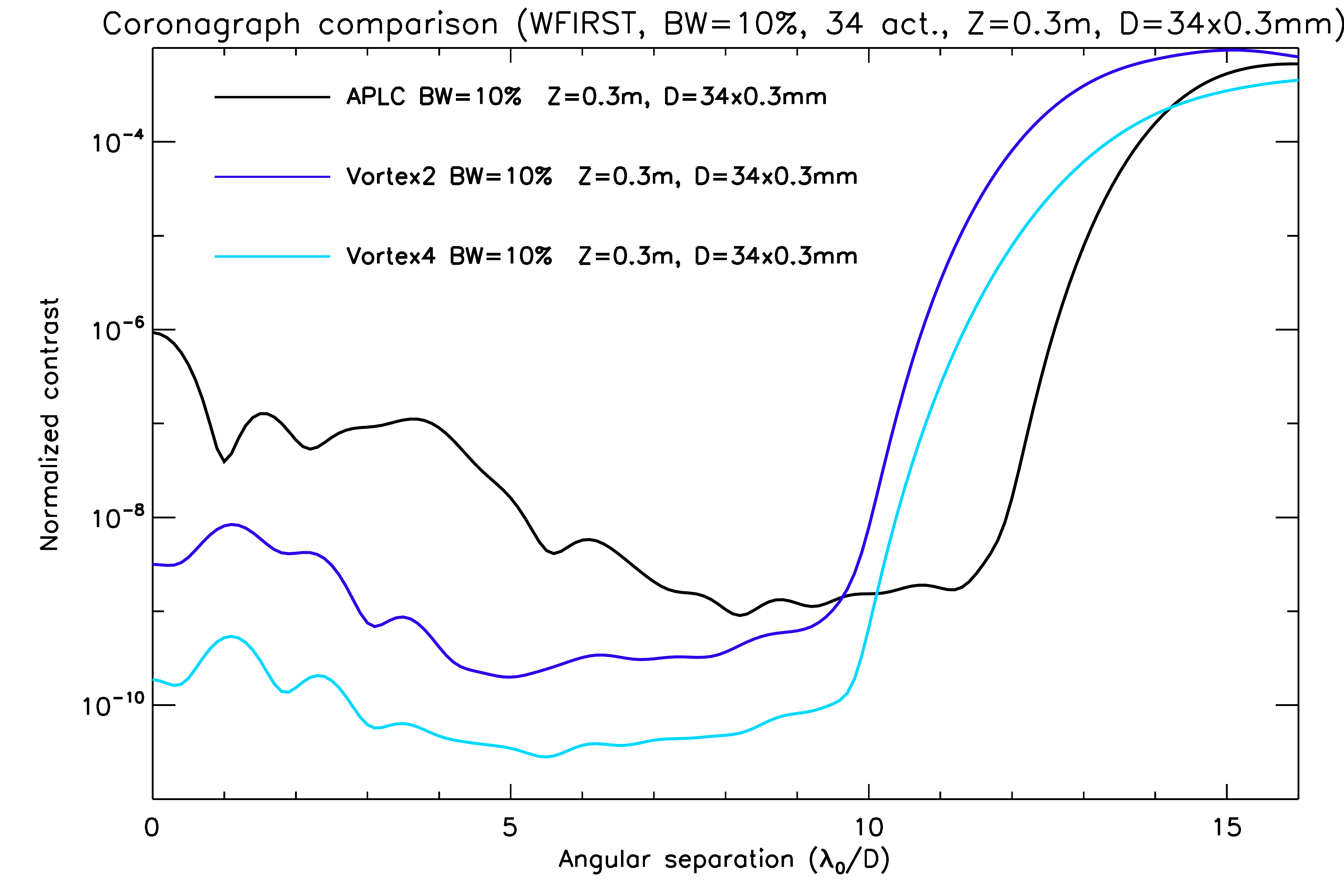}
    \includegraphics[width = 0.495\textwidth, trim= 0.5cm 0cm 0.3cm 0cm, clip = true]{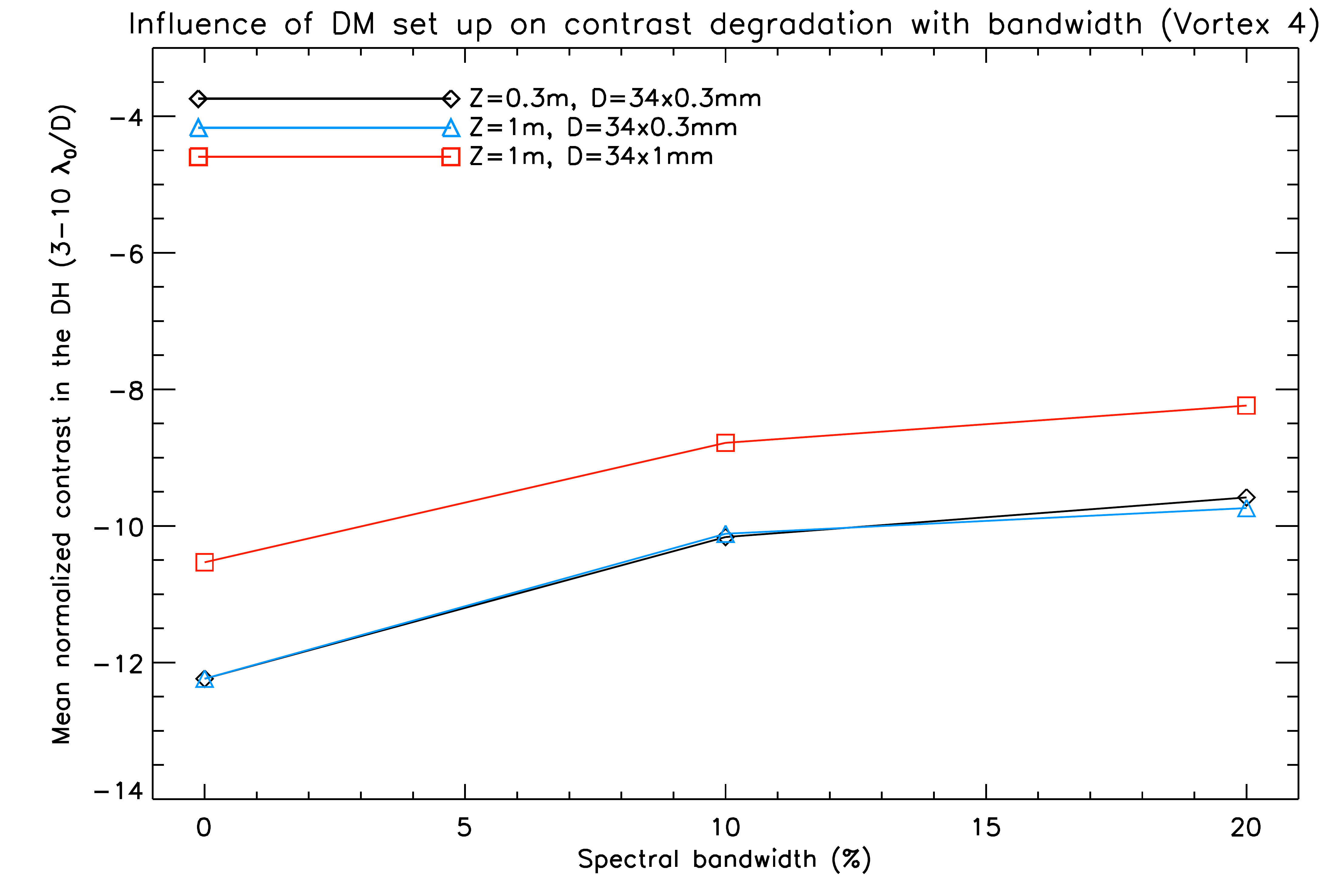}
\end{center}
 \caption[] 
{ \label{fig:WFIRST_contrast} 
\textbf{Results in contrast obtained with ACAD on the WFIRST pupil}. Left: Contrast curves for the 3 coronagraphs, with a 10\% bandwidth, 34 actuators, a DM size of 34x0.3mm, separated by 30cm. The dark hole is 5-12 $\lambda$/D for the APLC and 3 -10 $\lambda$/D for the Vortex coronagraphs. Right: Mean contrast in the DH for the charge 4 vortex, for 3 bandwidths (0\%, 10\%, 20\%) and three DM setups: the "HiCAT case" (small DMs, small inter-DM distance, black line), the "WFIRST case" (large DMs, long inter-DM distance, red line), and finally the "intermediate case" (small DMs, long inter-DM distance, blue line).}
\end{figure}

Figure~\ref{fig:WFIRST_pup_foc} presents some of the DHs obtained with the WFIRST pupil and the associated DM surfaces to provide them. The presented results were obtained with ACAD on the WFIRST pupil, for the APLC (4 images on the left) and charge 4 vortex coronagraph (4 images on the right), with a 10\% bandwidth. The DMs have 34 actuators, a DM size of 34x0.3mm, and are separated by 30cm ("HiCAT case"). For each coronagraph, we represent: the initial pupil (top left), the final DH obtained at the end of the correction (top right), 5-12 $\lambda$/D for APLC and 3 -10 $\lambda$/D for the Vortex, and the two shapes applied on the DMs to produce this DH (images at the bottom). We plotted the contrast curves for these results in Figure~\ref{fig:WFIRST_contrast} (left). As expected\cite{Mawet13}, the charge 4 vortex has a better contrast than the charge 2 vortex coronagraph. The contrast performance with APLC coronagraph is not as good but with only 31 actuators in the pupil, the outer working angle in this case (12 $\lambda$/D) is close to the actual DM limit (15.5 $\lambda$/D). In addition, this coronagraph was optimized for a contrast of $10^{-9}$ only, which means that we actually reached the performance obtained with this coronagraph on a continuous aperture, which is the goal of the ACAD technique. We are confident that with a slightly different APLC and a higher number of actuators, performance in contrast level could reach $10^{-10}$. On the other hand, it is unlikely that we manage to reduce the IWA under 5 $\lambda$/D for this type of coronagraph for an aperture with a 36\% central obstruction.

Figure~\ref{fig:WFIRST_contrast} (right) shows the influence of the DM setup on the contrast for the charge 4 vortex. The mean contrast in the DH is given for 3 bandwidths (0\%, 10\%, 20\%) and three DM setups: the "HiCAT case" (small DMs, small inter-DM distance, black line), the "WFIRST case" (large DMs, long inter-DM distance, red line), and finally the "intermediate case" (small DMs, long inter-DM distance, blue line). We clearly see that in term of contrast level, small DMs are to be favored. The inter-DM distance has no influence, except maybe at large bandwidths. Once again, this contrast levels are to be analysed in the context of this parametric study and should not be compared to the one of the WFIRST mission, as we only used 34 actuators in the pupil, and not 48 actuators. 

\subsection{Throughput and robustness to Jitter}
\label{sec:Throughput}

The throughput is measured as the "PSF core throughput" in Krist et al. 2016\cite{Krist16} (energy higher than half of the maximum in the final focal plane divided by the energy hitting the primary mirror). This measurement therefore includes the loss of energy in the PSF due to the coronagraph and also due to the deformation of the PSF because of the propagation between the 2 DMs. We normally expect that the deformation is more important (and therefore the throughput lower) when the strokes are important. 

 \begin{figure}[ht]
 	\parbox{0.5\textwidth}{ 
 	\centering \includegraphics[width = 0.5\textwidth, trim= 1.5cm 0cm 1.6cm 0cm, clip = true]{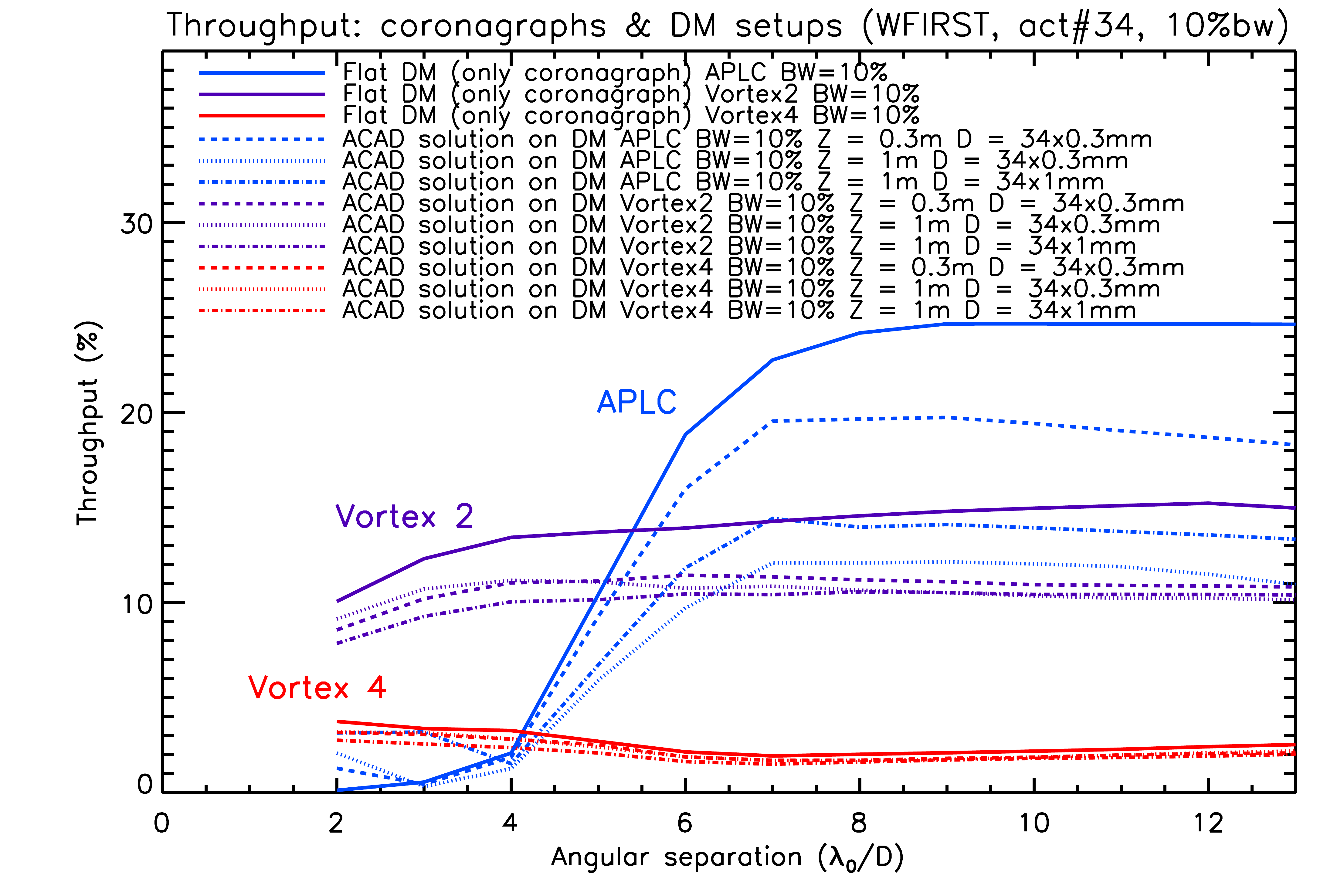}
 	}
 	\parbox{0.5\textwidth}{ 
 	\centering
    \begin{tabular}{|c|l|l|l|}
    \hline
    \multicolumn{1}{|c|}{\textbf{Strokes (nm)}}& \multicolumn{1}{c|}{APLC}& \multicolumn{1}{c|}{Vortex 2}& \multicolumn{1}{c|}{Vortex 4}\\ \hline
    \begin{tabular}[c]{@{}c@{}}Z=0.3m,\\D=34x0.3mm\end{tabular}      & \begin{tabular}[c]{@{}l@{}}DM1:168 \\ DM2:137 \end{tabular} & \begin{tabular}[c]{@{}l@{}}DM1:163 \\ DM2:155\end{tabular} & \begin{tabular}[c]{@{}l@{}}DM1:132\\ DM2:154\end{tabular} \\ \hline
    \begin{tabular}[c]{@{}c@{}}Z=1m,\\D=34x0.3mm\end{tabular} & \begin{tabular}[c]{@{}l@{}}DM1:296\\ DM2:367\end{tabular} & \begin{tabular}[c]{@{}l@{}}DM1:205\\ DM2:165\end{tabular} & \begin{tabular}[c]{@{}l@{}}DM1:104\\ DM2:107\end{tabular} \\ \hline
    \begin{tabular}[c]{@{}c@{}}Z=1m,\\D=34x1mm\end{tabular}         & \begin{tabular}[c]{@{}l@{}}DM1:273\\ DM2:317\end{tabular} & \begin{tabular}[c]{@{}l@{}}DM1:268\\ DM2:306\end{tabular} & \begin{tabular}[c]{@{}l@{}}DM1:206\\ DM2:198\end{tabular} \\ \hline
    \end{tabular}
}
\caption[] { \label{fig:WFIRST_throughput_stroke} \textbf{Throughput \& Strokes: WFIRST pupil, 10\% bandwidth} Left: Throughput as a function of separation (in $\lambda_0/D$) for 3 different coronagraphs and 3 DM setups. In solid line, we represent the throughput in the case of flat DMs (before the correction), which corresponds to the throughput of the coronagraph itself. In dashed, dotted and dash-dotted lines, we represent the throughput for the different DM setups, for APLC (blue), Vortex charge 2 (purple), Vortex charge 4 (red). Right: Strokes (in nm) for each DM, for the 3 coronagraphs and the 3 DM setups.}
\end{figure}

Throughput results are shown on Figure~\ref{fig:WFIRST_throughput_stroke} in the 10\% bandwidth case for the WFIRST pupil. We show the throughput as a function of separation (in $\lambda_0/D$) for 3 different coronagraphs and 3 DM setups. In solid line, we represent the throughput in the case of flat DMs (before the correction), which corresponds to the throughput of the coronagraph itself. In dashed and dotted lines, we represent the throughput for the different DM setups, for APLC (red), Vortex charge 2 (purple), Vortex charge 4 (blue). The most important parameter is the type of coronagraph. The ring-apodized vortex (especially the charge 4) is not at all optimized for a pupil this size. As shown in Table\ref{tab:corono}, in the case of the ring apodized charge 4 coronagraph, the central apodization of the Lyot stop is of 78\%, blocking most of the energy. Other methods of apodization for the charge 4 vortex (e.g. apodized phase mask + vortex\cite{Carlotti14}) are currently investigated to enhance the throughput. In the same figure, we can also compare the influence set up of the throughput of the system. For charge 4 vortex, throughput is mostly the same in every case, which we interpret by the fact that the off-axis PSF is already completely deformed by the coronagraph itself. For the APLC, the "HiCAT case" (small DMs, small inter-DM distance, dashed line) has the most throughput, followed by the "WFIRST case" (large DMs, long inter-DM distance, dot-dashed line), and finally by the "intermediate case" (small DMs, long inter-DM distance, dotted line). Unsurprisingly, this ranking is also the one followed by the strokes (table on the left, first column). For the charge 2 vortex the "HiCAT case" (small DMs, small inter-DM distance, dashed line) has the most throughput, but the two other cases ("WFIRST case" and "intermediate case") are mostly identical. More analysis (5 or 6 cases for both DM size and inter-DM distance) are necessary to fully understand the influence of the DM set-up on the throughput.

A quick analysis of the robustness to jitter shows that the type of coronagraph is the most important parameter for this (with little of no influence of the DM setup or the bandwidth). As expected, the APLC is the most jitter-resistant coronagraph (basically no change in contrast in the DH before $3.10^{-2} \lambda/D$), followed by charge 4 vortex and finally charge 2 vortex (degradation in contrast in the DH as soon as $1.10^{-3} \lambda/D$).

\section{Performance on one of the SCDA pupil}
\label{sec:SCDA}

\begin{figure}[t]
 \begin{center}
    \includegraphics[width = 0.49\textwidth, trim= 0cm 4.5cm 5cm 4cm, clip = true]{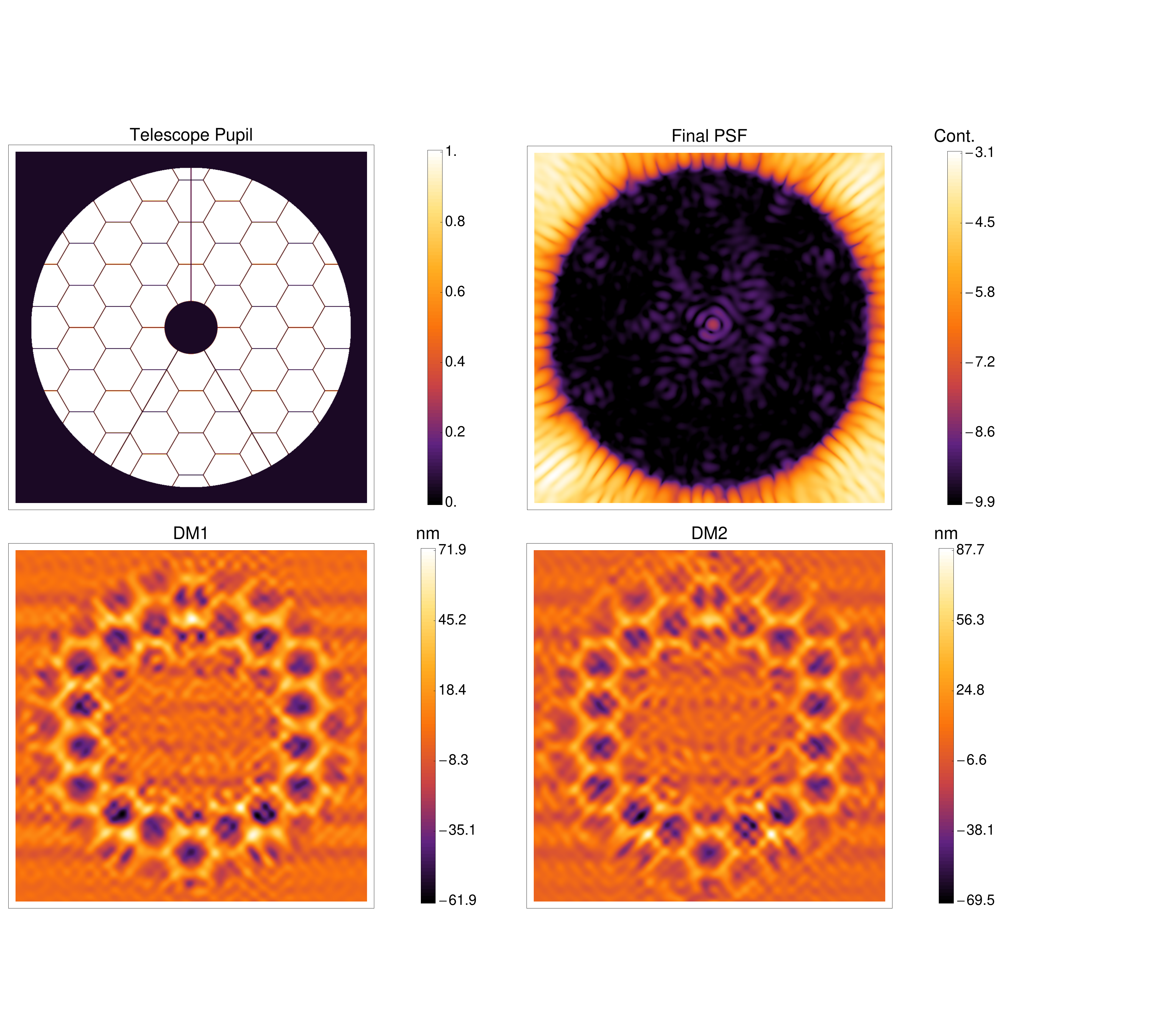}
    \includegraphics[width = 0.49\textwidth, trim= 0cm 0cm 1cm 0cm, clip = true]{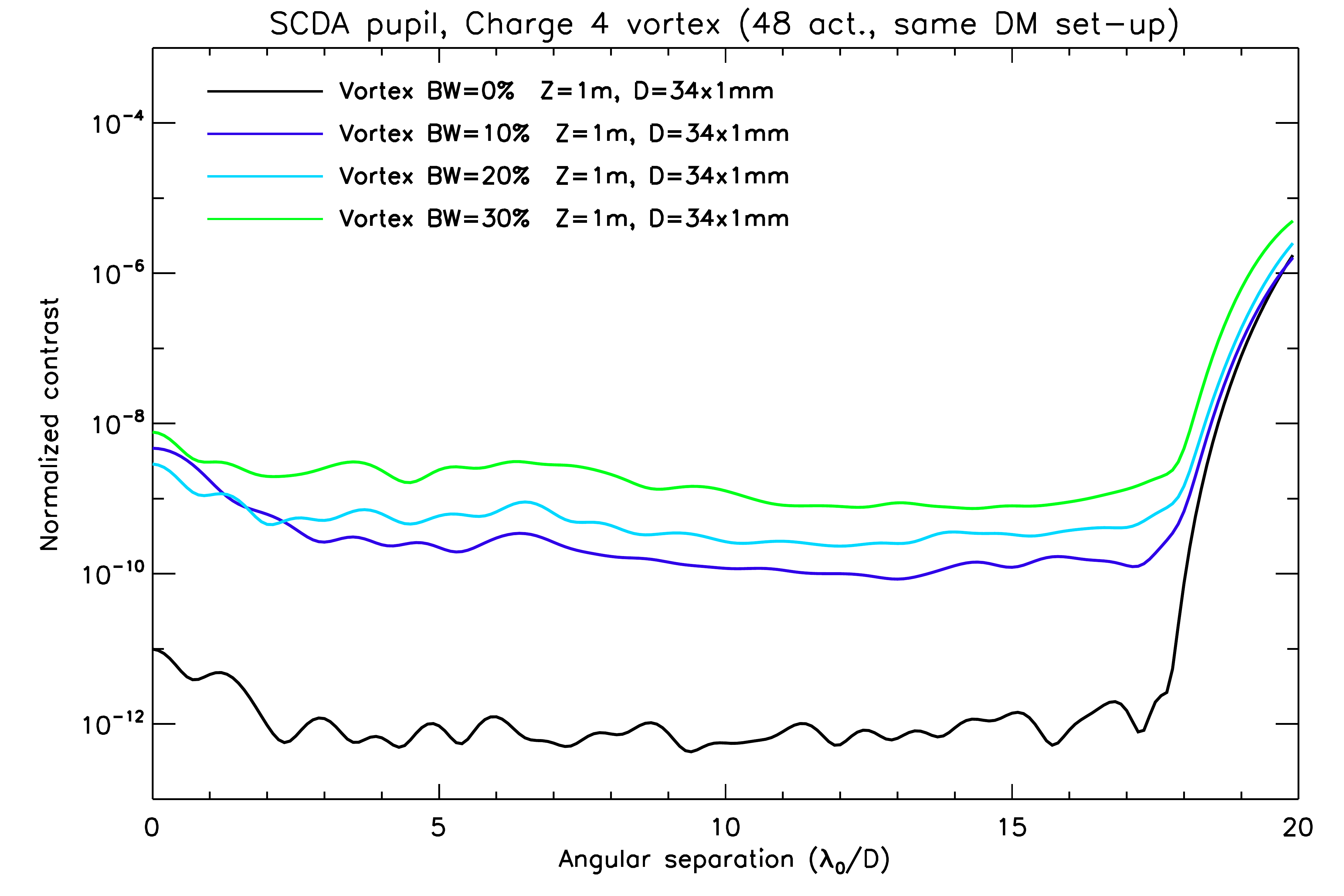}
\end{center}
 \caption[] 
{ \label{fig:SCDA_contrast} 
\textbf{Results obtain with ACAD on the SCDA pupil, for the ring apodized Charge 4 vortex.} The DM have 48 actuators, and is "Xinetics" like : the inter-actuator distance is 1 mm and the DM is therefore 4.8 cm. The inter-DM distance is 1 m. On the left, we present the initial pupil (top left), the final 1.5-18 $\lambda_0$/D DH obtained at the end of the correction (top middle) and the two shapes applied on the DMs to produce this DH (bottom left and middle). On the right we present the azimuthal profile of the contrast level in the DH as a function of the distance to the star in $\lambda_0$/D}
\end{figure}

In this Section, we wanted to present the first results obtain on a realist aperture for a large space based mission (See Figure~\ref{fig:SCDA_contrast}, top left). This aperture is part of the apertures of the Segmented Coronagraph Design and Analysis (SCDA\footnote{See http://exep.jpl.nasa.gov/files/exep/SCDAApertureDocument0504161.pdf}) task, launch by Exoplanet Exploration Program (ExEP) Office to determine scientific performance of candidate coronagraph instruments for large space telescopes with segmented, obstructed apertures. Compare to WFIRST apertures, the selected aperture is segmented but the have a central obstruction of 17\% only and finer struts, which make it far easier.

\begin{table}[ht]
\centering
\caption{Stroke after SM for SCDA pupil and several bandwidth}
\label{tab:strokeSCDA}
\begin{tabular}{|l|l|l|l|l|}
\hline
                                                                                                       & \multicolumn{1}{c|}{0\% bandwidth}                                & \multicolumn{1}{c|}{10\% bandwidth}                               & \multicolumn{1}{c|}{20\% bandwidth}                               & \multicolumn{1}{c|}{30\% bandwidth}                               \\ \hline
\multicolumn{1}{|c|}{\begin{tabular}[c]{@{}c@{}}Vortex (charge 4)\\ D = 48x1mm, Z = 0.3m\end{tabular}} & \begin{tabular}[c]{@{}l@{}}DM1: 102 nm\\ DM2: 110 nm\end{tabular} & \begin{tabular}[c]{@{}l@{}}DM1: 125 nm\\ DM2: 142 nm\end{tabular} & \begin{tabular}[c]{@{}l@{}}DM1: 125 nm\\ DM2: 143 nm\end{tabular} & \begin{tabular}[c]{@{}l@{}}DM1: 113 nm\\ DM2: 133 nm\end{tabular} \\ \hline
\end{tabular}
\end{table}

This section is not really a parametric study, we only analyze a few situations to determine the performance of our technique with an aperture easier than WFIRST's one. We take the least favorable DM setup, of the 3 studied before: two "Xinetics" like DMs (1 mm inter-actuator distance) separated by 1 m. We use 48 actuators. This is therefore the exact setup of the WFIRST mission. We set the IWA to 1.5 $\lambda_0/D$ and the OWA to 18 $\lambda_0/D$ and run the method for several bandwidths (0\%, 10\%, 20\%, 30\%). 

\begin{figure}[ht]
 \begin{center}
    \includegraphics[width = 0.49\textwidth, trim= 0cm 0cm 1cm 0cm, clip = true]{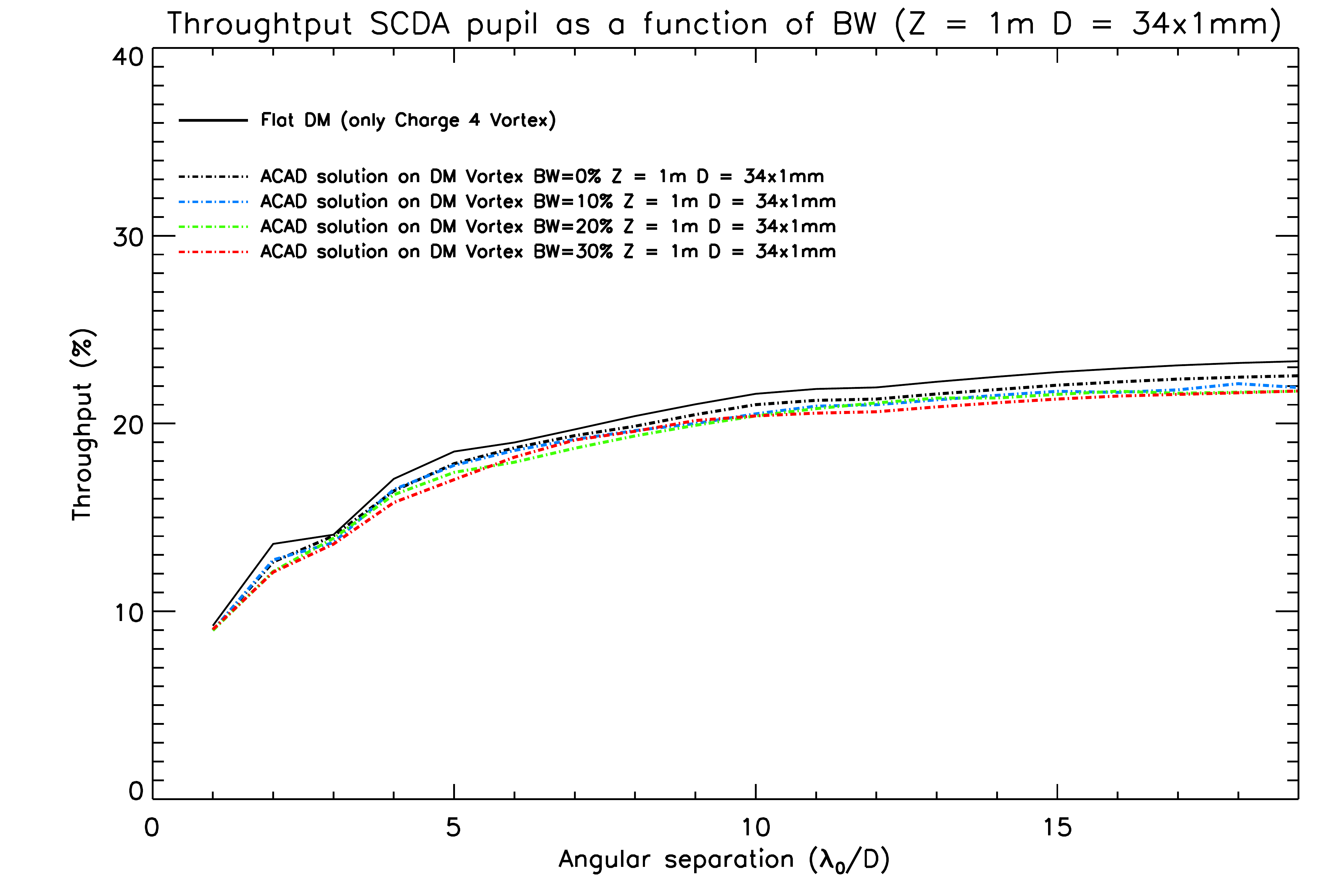}
    \includegraphics[width = 0.49\textwidth, trim= 0cm 0cm 1cm 0cm, clip = true]{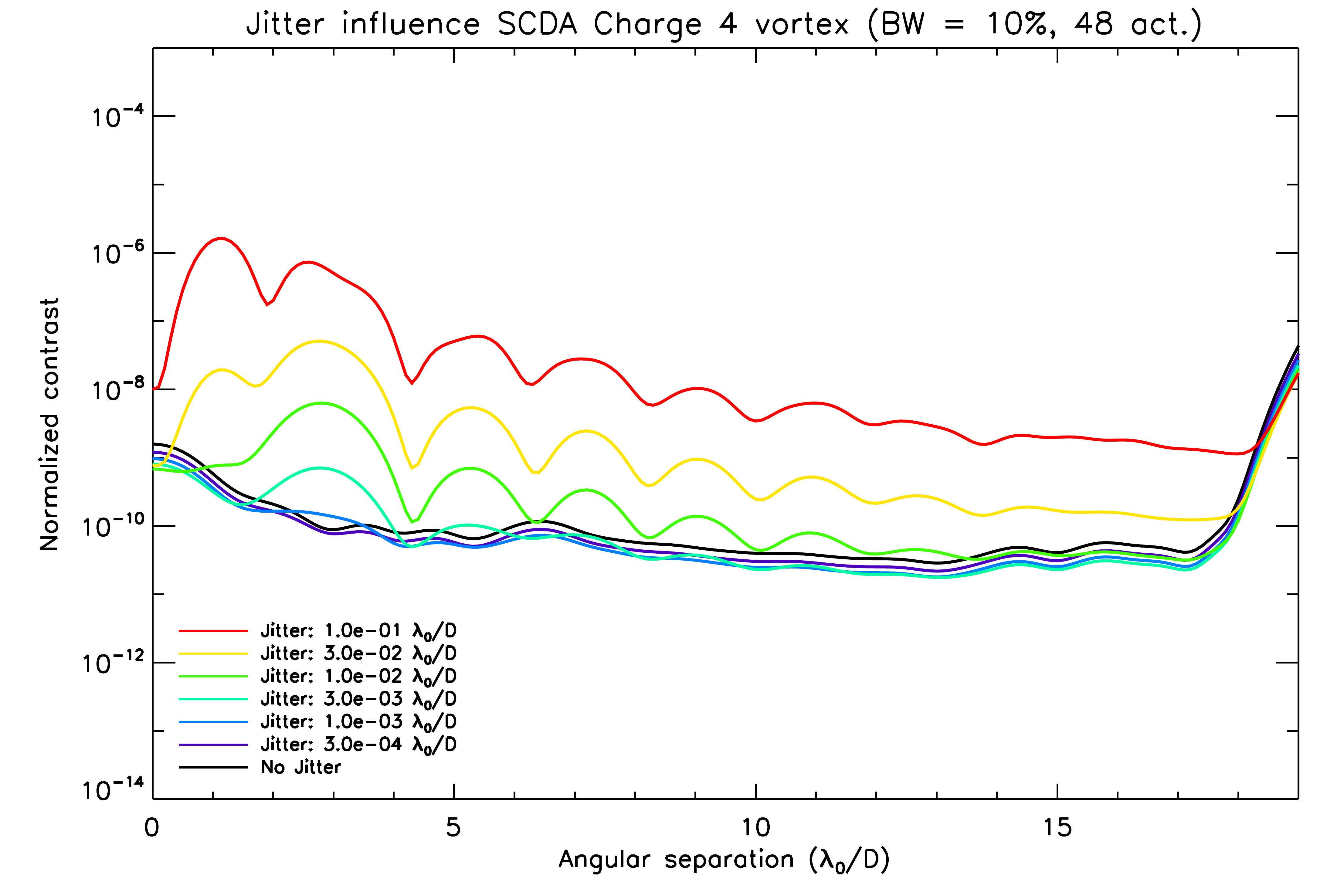}
\end{center}
 \caption[] 
{ \label{fig:SCDA_through_jitter} 
\textbf{Results obtained with ACAD on the SCDA pupil, for the Charge 4 vortex.} The DM have 48 actuators, and is "Xinetics" like : the inter actuator distance is 1mm and the DM is therefore 4.8 cm. The inter-DM distance is 1 m. Left: throughput over the DH, for each bandwidths (0\%, 10\%, 20\%, 30\%). Right: Influence of different levels of jitter on contrast performance on the DH for the 10\% bandwidth.}
\end{figure}

The results are presented in Figure~\ref{fig:SCDA_contrast} and \ref{fig:SCDA_through_jitter}. In Figure~\ref{fig:SCDA_contrast}, we present the results in contrast level. We show the initial SCDA pupil (top left), the final 1.5-18 $\lambda_0$/D DH obtained at the end of the correction (top, middle) and the two shapes applied on the DMs to produce this DH (bottom left and middle). On the right we present the azimuthal profil of the contrast level in the DH as a function of the distance to the star in $\lambda_0$/D. In term of contrast level, we obtain a mean of less than $2.10^-{10}$ over the 1.5-18 $\lambda_0/D$ DH and a 10\% bandwidth and a mean of less than $5.10^-{10}$ over the same DH a 20\% bandwidth. The advantage of the ACAD technique is its adaptability: if needed, we can improve contrast at the expense of OWA or spectral bandwidth.

The Throughput is presented on Figure~\ref{fig:SCDA_through_jitter} (left) for every bandwidth (0\% to 30\%). We obtain a throughput better than 10\% on the whole DH. As already notice in the previous section, the bandwidth of the correction have little influence on the throughput. Finally, the Robustness to Jitter is presented in Figure~\ref{fig:SCDA_through_jitter} (right). The influence of jitter on the contrast is limited for jitter lower than $1.10^{-2} \lambda_0/D$. For larger jitters, the contrast start to degrade in the DH. We checked that the jitter robustness was not impacted for different bandwidths. As expected, the throughput degrades as the strokes (presented in Table~\ref{tab:strokeSCDA}) increase.

\section{Conclusion}

The main result of this paper is that we finally obtained ACAD performance that can be used to design space-based instruments ($10^-{10}$ with a 10\% spectral bandwidth for the WFIRST aperture or one SCDA pupil). We can now start to implement ACAD method for SCDA task force apertures and compare its results to the other techniques currently under review.

The parametric study show several important results. First, as expected, the increase of spectral bandwidth has little impact on the throughput and robustness to jitter. Secondly and more importantly, it seems that in our current method, we cannot reduce the influence of set-up of the DMs to the value $D^2/Z$ like we did in previous papers. Indeed, the two parameters seems to have independent impact on different metrics. On the one hand, the inter-DM distance seems to have a strong impact on the throughput but not on the contrast. On the other hand, the size of the DMs seems to have a important impact on both the contrast level and throughput. We need to expand the analyze for several more cases for D and Z to fully decorelate the effects of these two parameters on the metrics. Eventually if we can confirm that the size of the DMs is such an important parameter, this could means that there is an optimal number of actuator above which the DM size would so large it would have a negative impact on contrast. 

It is also primordial to understand the importance of ray optic solution (given by the numerical resolution of the Monge-Amp\`ere equation) in the context of high contrast correction. This paper proves that there are several local minima in contrast achievable "linearly" by the SM algorithm, starting from given initial shapes. We discovered that the local minimum around the ray optic solution, although close or identical in contrast level, requires higher strokes, which in turn degrades throughput. A first explanation is that, as repeatably noticed by many people in this field \cite{Mazoyer14}, flattening the wavefront at the entrance of the coronagraph or digging high contrast zones in the focal plane are two very different things. However, there might be other advantages if we use the ray optic solution first step, for large bandwidths correction. 
Finally, the next application will be to put this technique to the experimental test. The High-contrast imager for Complex Aperture Telescopes (HiCAT) bench\cite{NDiayeSPIE15} is currently under development and should soon obtained its first DH with multiple DMs. Then ACAD experimental validation should follow quickly. Simulation results in this paper and in Mazoyer et al. 2016\cite{Mazoyer_JATIS} show that the ACAD shapes are achievable with the HiCAT bench's DMs. The new result on this paper that the ray optic solution step may not be necessary will make the experimental test a lot easier, because we do not need open loop control of the DMs anymore.

\acknowledgments     
 
This material is based upon work carried out under subcontract \#1496556 with the Jet Propulsion Laboratory funded by NASA and administered by the California Institute of Technology. We would like to thanks J. Krist (JPL) for his helpful comments on the throughput and contrast definition.


\bibliography{biblio_spie_edh16}   
\bibliographystyle{spiebib}   

\end{document}